# Modeling of biological doses and mechanical effects on bone transduction

Romain Rieger, Ridha Hambli* and Rachid Jennane

PRISME Laboratory, University of Orleans

8 rue Léonard de Vinci, 45072 Orléans cedex 2, France

Phone : +33 (0)238 49 40 55

Fax : +33 (0)238 41 73 83

* Corresponding author: ridha.hambli@univ-orleans.fr

**Abstract**

Shear stress, hormones like parathyroid and mineral elements like calcium mediate the amplitude of stimulus signal which affects the rate of bone remodeling. The current study investigates the theoretical effects of different metabolic doses in stimulus signal level on bone. The model was built considering the osteocyte as the sensing center mediated by coupled mechanical shear stress and some biological factors. The proposed enhanced model was developed based on previously published works dealing with different aspects of bone transduction. It describes the effects of physiological doses variations of Calcium, Parathyroid Hormone, Nitric Oxide and Prostaglandin E2 on the stimulus level sensed by osteocytes in response to applied shear stress generated by interstitial fluid flow. We retained the metabolic factors (Parathyroid Hormone, Nitric Oxide, and Prostaglandin E2) as parameters of bone cell mechanosensitivity because stimulation/inhibition of induced pathways stimulates osteogenic response in vivo. We then tested the model response in term of stimulus signal variation versus the biological factors doses to external mechanical stimuli. Despite the limitations of the model, it is consistent and has physiological bases. Biological inputs are histologically measurable. This makes the model amenable to experimental verification.

**Keywords:** Osteocytes; Shear stress; Ca-PTH; NO; $PGE_2$

**Notations**

$x_{ocy}$ : Active sensing osteocytes number.

$x_b$ : Active osteocblasts number.

$x_c$ : Active osteoclasts number.

$x_{Ca}$ : Calcium level

$x_{PTH}$ : PTH level

$x_{NO}$ : Nitric Oxide level.

$x_{PGE}$ : Prostaglandin E2 level.

$V_P$ : Interstitial fluid velocity generated by pressure.

$\dfrac{dP}{dz}$ : Pressure gradient in the canaliculi; $z$ denotes the axial coordinate of the canaliculi.

**1. Introduction**

It is well admitted that mechanical strain is one of the main stimulus triggering bone remodeling.

Bone adaptation to its environment is influenced by both mechanical and biological stimuli (Feskanich et al., 2003; Goltzman, 1999; Heldring et al., 2007). Fluid flow imposes a shear stress on osteocytes that appears to deform the cells (Weinbaum et al., 1994).

Bonewald (2008) rewieved a cascades of transduction events into the osteocyte as a response to mechanical loading. Important signaling molecules like Nitric Oxide (NO) and Prostaglandin E2 ($PGE_2$) influencing bone remodeling homeostasis have been shown to be produced in the osteocyte in response to fluid flow.

Weinbaum et al., (1994) were one of the first to propose a theoretical model to predict the fluid shear stress and streaming potential at the surface of osteocytic process in the lacunocanalicular porosity. Rémond et al., (2005) enhanced the model of Zeng by including mass flux. Lemaire et al., (2005) developed a more complex model based on the idea that the motion of interstitial fluid is caused by a combination of mechanical strain, electro-osmotic and osmotic actions.

Few numerical studies modeling the load-induced fluid flow have been developed (Goulet et al., 2008; Gururaja et al., 2005; Roland Steck, 2003; Swan et al., 2003). Some other theoretical and numerical studies have been performed considering the osteocyte's tissue strain amplification (Han et al., 2004; Rathbonivtch et al., 2007) or the integrin function on the mechanosensation process (Wang et al., 2007; Weinbaum, 2003).

Recently, Adachi et al., (2010) developed a bone remodeling finite element model based on a stimulus function expressed in term of fluid flow in the lacunocanalicular system. The limitations of this work it that the model neglects the coupling effects with the main biological mechanisms. No biological entities were considered and the bone adaptation is purely triggered by mechanical stimulus.

Some studies have attempted to explore the effect of metabolic doses regarding bone remodeling or bone diseases in numerical models. Maldonado, (2006) has developed a formulation of the production of NO and $PGE_2$ by the osteocyte and a description of the evolution of the osteocyte cell population. The osteocyte rate growth was expressed as a function of the number of osteoblasts (some osteoblasts differentiate into osteocytes) and a fluid shear stress function (mechanical loading is involving on osteocyte viability).

Recently, enhanced mathematical model was developed by Peterson and Riggs (2010) to describe the bone and Calcium-Parathyroid Hormone (Ca-PTH) homeostasis. Komarova (2005) have developed a theoretical model describing the autocrine and paracrine factors which are involved in the Bone Multicellular Units (BMUs) auto-regulation composed of osteoblast and osteoclast. In this work the authors studied the influence of pulsatile PTH therapy on those autocrine and paracrine factors to induce bone formation. Lemaire et al., (2004) have developed a more dynamic BMUs interaction model and investigated the effects on bone remodeling of the different biological factor such as Osteoprotegerin (OPG) or Transforming Growth Factor β (TGF-β).

Some others have investigated the effect of mechanical stimulus on bone remodeling without considering the metabolic factor effects (Hambli et al., 2010; Hambli et al., 2009; McNamara and Prendergast, 2007; Scott et al., 2001).

Despite of the large number of studies dealing with bone mechanotransduction, there is still a lack of mechanobiological models combining in a unified way the most observed mechanical and biological mechanisms during transduction phase on bone (Rieger et al., 2010). Such models are very useful to enhance bone remodeling predictions based on "biophysical" description rather than phenomenological ones.

Oesterhelt (2010), proposed a general framework to describe the transduction concept based on four steps. (i) Reception of the external signals as inputs (mechanical and biological). Signals can be stimulatory or inhibitory; (ii) Integration (combination) of the signals (iii); Amplification of the signals as outputs (iv); Signals sent to the specific target.

Following the transduction concept proposed by Oesterhelt (2010) the current study attempts to investigate the effect of doses in transduction signal based on a plausible theoretical model.

Several physiological factors can mediate the transduction. In the present work, we retained Ca, PTH, NO and $PGE_2$ factors as parameters of the model because those metabolic factors are known to stimulate/inhibit osteogenic response in vivo.

## 2. Materials and methods

### 2.1. Concepts of bone transduction

A summary of the previous discussed components is presented in table 1. The integrin senses the amount of Ca and regulates PTH. In the same manner the integrin produces NO, $PGE_2$ and a mechanical stimulus due to fluid flow. All those components are organized into mechanical and biological normalized signals and then summed up into a full mechanobiological stimulus.

Table 1

### 2.2. Proposition of a transduction model

Based on Oesterhelt (2010) transduction concept combined with Bonewald et al., (2008) we propose the following general block diagram representing the osteocyte transduction phase (Fig. 1):

Figure 1

## 2.3. Mathematical description

The proposed model is based on the idea that the osteocyte is the operating center which generates the interaction between both mechanical stimulus and biological factors. After combining both stimuli (mechanical and biological), the osteocyte triggers the signal to active BMUs to form or resorb bone.

For the signaling phase we consider the fluid flow velocity in the canaliculi due to mechanical pressure proposed by Lemaire T. et al., (2005). They concluded that the fluid movement $V_P$ (equ. (1)) generated by pressure gradient is the most important driving effect in fluid flow since it accounts for 95%. The term $(1-D)^\gamma$ was added to Darcy's law to incorporate the bone matrix damage ($D$) where $\gamma$ is a damage exponent that defines the sensitivity of the damage in the fluid velocity.

The biological signaling part is assumed to be triggered by Ca demands. The Ca level ($x_{Ca}$) is considered as an input in the model (equ. (2) which leads to PTH release ($x_{PTH}$). To describe the mathematical relation between PTH release rate and level of Ca, we propose a fitted model obtained by experiments (Haden et al., 2000; Houillier, 2009) expressed by equation (3) exhibiting a sigmoidale relation.

The reception phase is expressed by the fluid velocity (equ. (4)) where $V_P = V_T$ and the release of NO ($x_{NO}$) and PGE$_2$ ($x_{PGE2}$) (equ. (5-6)), from the integrin.

The NO and PGE$_2$ rates growth (Maldonado, 2006) depends on a fluid shear stress function $f_{V_P}$ (equ. 7) applied to osteocytes. If NO evolution is purely stimulated by the fluid shear stress function, PGE$_2$ are also dependant of the NO evolution and then the coupling is expressed in equation (6).

Osteocyte detects the concentration of Ca through PTH (Langub et al., 2001; Teti and Zallone, 2009) and responses to the fluid shear stress. Thus the Osteocyte population growth ($x_{ocy}$) developed by Maldonado et al., (2006) has been enhanced and is expressed by equation (8). It is directly affected by the osteoblasts population ($x_b$) and the osteocytes.

Komarova et al., 2003 developed a model describing the BMUs population dynamics (equ. 9, 10). It includes osteoblasts ($x_b$), osteoclasts ($x_c$) and considers an auto-regulation between these two types of cells. Indeed the autocrine and paracrine factors are phenomenological modeled as a rate coefficient by an exponent in the model. Insulin-like Growth Factor (IGF) (g22) and Transforming Growth factor Beta (TGF-β) (g11) are autocrine factor stimulating osteoblasts and osteoclasts respectively. Whereas for the stimulation of osteoblasts the paracrine factor is modeled by an exponent (g12) representing IGF and TGF-β actions. Regarding osteoclasts the paracrine stimulation takes effects trough the exponent (g21) modeling Receptor Activator of Nuclear factor κB Ligand (RANKL) and Osteoprotegerin (OPG) actions.

Ryser et al., (2009) developed a model describing the evolution of a single BMU taking into account dynamic biochemical factors such as RANKL and OPG driving osteoblasts and osteoclasts evolution. Moreover it describes temporal and spatial features of the digging hole and the movement of the BMUs across the bone surface. This model is biochemically more complex than the one from Komarova et al., (2003) and is more suited for visual bone remodeling evolution.

In order to investigate the potential in our approach to model the coupling effects on the transduction signal, the model of Komarova (2003) has been retained here for its simplicity in the autocrine and paracrine factors description which allows us to focus on the effects of the transduction mechanism.

The integration and amplification phase ensures a normalized signals variations between 0% and 100% of their maximum expressed by the relation $x_i^n = x_i/x_i^{Max}$, (equ. 11-14). It enables to leave aside dimensional consideration of the variables since we combine mechanical dimension (MPa) and biological concentration (mM) (Brazel and Peppas, 1999; Coatanéa E; Vareille J., 2003).

Finally, the mechanical and the biological signals are weighted and summed up (equ. 15-17) to generate a normalized stimulus function as a signal (varying between 0% and 100%) mediated by the biological factors (equ. 17).

$W_i$, i= Mecha, Bio, MB, PTH, NO, PGE$_2$: are mechanical and biological weight factors verifying: $0 < W_i \leq 1$ and $W_{Mecha} + W_{Bio} = W_{PTH} + W_{MB} = W_{NO} + W_{PGE2} = 1$.

Table 2

Table 3

## 4. Results

We have modeled and investigated the main important phenomena of bone transduction in a mechanical and biological point of view. The influence of metabolic doses on stimulus level has been explored in order to investigate their sensitivity on the mechanobiological stimulus. The following figures show the ability of the model to interpret mechanical effects due to bony fluid flow through the lacunocanalicular system and/or different biological effects due to the virtual injection or the basal production of different messengers such as NO and $PGE_2$.

If mechanical effect is the most influenced part in bone remodeling it is also highly dependent on the biological components considered as a regulating factor. Figure 2 shows the evolution of the stimulus versus of the Ca concentration for different magnitudes of the fluid flow.

Figure 2

Two observations should be outlined:

(i) For low quantity of Ca the model predicts an important secretion of PTH in order to resorb bone which results to an increase in Ca concentration. On the opposite, high values of calcium lead to a decrease of PTH secretion which results in an elevation of bone formation to stock free Ca into the bone. Those results are consistent with the observed biological reactions of Ca metabolism in bone emphasized by Parfitt (1976) and Houillier (2009) which tends to maintain a constant calcemia.

(ii) The difference in magnitude between the three curves reflects the impact of the fluid flow in the mechanobiological stimulus. Stronger gets the fluid flow, higher the stimulus becomes. Experimental issues from Burger et al., (1999) and Klein-Nulend (2005) confirm the predictive results of the model. In function of the level of the mechanobiological stimulus the transduction process favors bone's resorption or formation.

The mechanobiologic stimulus is directly dependent on the production of NO and PGE$_2$ by the osteocyte through the integrin stimulated by the fluid flow. The model can consider physiological evolution of such messenger or can be shunted by the consideration of some virtual constant injection. Figure 3 shows the prediction of the stimulus versus fluid velocity for different evolutions of NO and PGE$_2$.

Figure 3

It is well known that NO and PGE$_2$ have an important role in bone remodeling regulation (Brandi et al., 1995; Pilbeam et al., 2008; Wimalawansa, 2008). One can notice that fluid flow gradually increases the production of NO and PGE$_2$, which results in a higher stimulus. Although this stimulus can be increased or decreased at some specific magnitude of fluid flow by considering a virtual injection of continuous controlled dose (0.37 & 0.2 for the square curve; 1.6 & 0.2 for the circle curve; 1.6 & 0.85 for the triangle curve respectively for NO and PGE$_2$ [pM]) in a specific area. By doing so the model is able to magnify or to reduce NO and PGE$_2$ effects in the integral domain of influence. The model is considered as dose dependent because it enables the stimulus to be adjusted by the amount of NO and PGE$_2$ injected or auto regulated.

Since this is an original work, there is no comparable study which allows us to calibrate the appropriate values of the weighting-factors. Thus we propose two different configurations of the stimulus equations which represent the possibility of the model to match with the large adaptability of any biological system. Figure 4 exhibits the stimulus' responses to different combinations of weighting-factors. In fact we propose in the first case to express the stimulus to be equally responsive to the mechanical and biological part. This is expressed by assigning a weighting-factors of 0.5 (filled solid curves). In the second case the stimulus is more influenced by the biological component which is expressed by weighting $f_{Bio}$ by 0.8 (open dashed curves). Moreover for the second case we consider the effect of NO and PGE$_2$ to be more influencing than the PTH, which is represented by weighting $f_{MB}$ by 0.8.

Figure 4

For an equally distributed influence of the mechanical and biological components (solid filled solid curves), the stimulus responses gradually as the amount of NO [pM] and PGE$_2$ [pM] increases. Also the shape of the plot represents the evolution of PTH [pg.ml$^{-1}$] in function of Ca [mM] concentration. It should be noted that the shape of the open dashed curves is diminished which is explained by a decrease in the influence of PTH in the mechanobiological stimulus.

Our model incorporates the effects of the bone matrix damage. Fluid velocity is obtained by Darcy's law and is weighted by the term $(1-D)^{\gamma}$. The damage considered here is due to the presence of microcracks into the bone which leads to an alteration of the osteocyte's communication (Burger and Klein-Nulend, 1999; Prendergast and Huiskes, 1996). Instead of an open/close communication relation we shall consider here the possibility of some degradation of the signals sent trough the lacunocanalicular network. So when the damage increases it reduces the magnitude of the mechanical signal sent to the osteocytes. The figure 5 depicts the influence of the level of damage in the mechanobiological stimulus for basal physiological evolution of NO and PGE$_2$.

Figure 5

The disrupted communication between osteocytes increases as the damage on the lacunocanalicular network increases. This can be represented by a leakage of the bony fluid contained in the canaliculi. So the proportionality between the fluid flow and the deformation applied onto the bone is altered and the resulted signal does not match properly with the expected information. This non-proportionality is expressed by the modification of the stimulus' magnitude.

## 5. Discussion

The work presented here is an attempt to develop a dose dependent integrated transduction model incorporating both mechanical and biological reactions. The idea has arisen from the observation that actual transduction model are only based on mechanical stress neglecting the biological effects.

The current model predicts the influence of Ca concentration on the magnitude of the mechanobiological stimulus. The concentration of Ca can be a regulating factor for the stimulus increase. Low level of Ca induces an important release of PTH (Houillier, 2009) which stimulates bone resorption in order to deliver Ca from the bone to restore falling plasma Ca. This effect leads to a reduced level of the stimulus. On the opposite high concentration of Ca diminishes the release of PTH and as a consequence, it stimulates bone formation. This is consistent with the work of Chen et al., (1999) who have shown that there was a progressive increase in the PTH secretion as dietary Ca level decreased which would favor Ca mobilization from bone.

Experimental (Bakker et al., 2001; Klein-Nulend et al., 2005; Pitsillides et al., 1995) and numerical (McGarry, 2004) studies have shown NO and $PGE_2$ to be increased when fluid flow grows which confirms the results of the model for two reasons. Firstly we can observe the increase of NO and $PGE_2$ as the fluid flow magnifies. Secondly the stimulus is amplified as the amount of NO and $PGE_2$ increase. Based on the work of Maldonado (2006) the kinetics of NO and $PGE_2$ have been integrated in the present study in order to create a more precise stimulus taking into account the influence of NO and $PGE_2$.

Other studies have proven the importance and the benefit of non-mechanical agent to enhance mechanical loading (Jee WS, 2005). This confirms the regulating role of biological messengers which inhibit the action of mechanical loading or improve bone adaptation responsiveness. Their action can be easily illustrated in some hormonal bone diseases by submitting patients onto drugs administration (mainly non-mechanical agents such as estrogen or growth hormone). The threshold values are then moved in order to improve mechanical loading effects.

Nevertheless, the model presents some limitations and simplifications. First, the mechanotransduction process is much more complex than the one described in this paper. Vitamin D (St-Arnaud, 2008; Wasserman et al., 1995), Ca homeostasis (Peterson and Riggs), age, sex, drugs treatments and nutrition should be taken under consideration in order to get an accurate model of transduction. Building such comprehensive model requires proceeding step by step and implies to develop a block diagram model in order to be easily updated.

Second, our model does not consider the pulsatile and the continuous injection of PTH. Many studies suggest an anabolic effect of PTH for pulsated injection (Komarova, 2005; Kousteni and Bilezikian, 2008) and catabolic one for continuous injection. An extended version of the model could incorporate the pulsatile/continuous PTH dosing (Potter et al., 2005).

Experiments are required to further study the many parameters required for the model and the effects of biological factors dosing the mechanobiological stimulus. Further investigation of the molecular basis of mechanotransduction on bone physiology and disease is needed to confirm that the mechanobiological model can provide realistic predictions.

In spite of these limitations the model can predicts a plausible mechanobiological stimulus consistent with experimental, clinical and numerical observations. Moreover it is easily updatable regarding the block diagram architecture.

In conclusion two perspectives arise from this work. First of all we are going to link this work with a model of BMUs in order to define a complete bone remodeling algorithm (Rieger et al., 2010). Then the whole model will be implemented into finite element code in order to predict bone quality evolution for specific patients and to estimate the necessity of putting them under drugs medication.

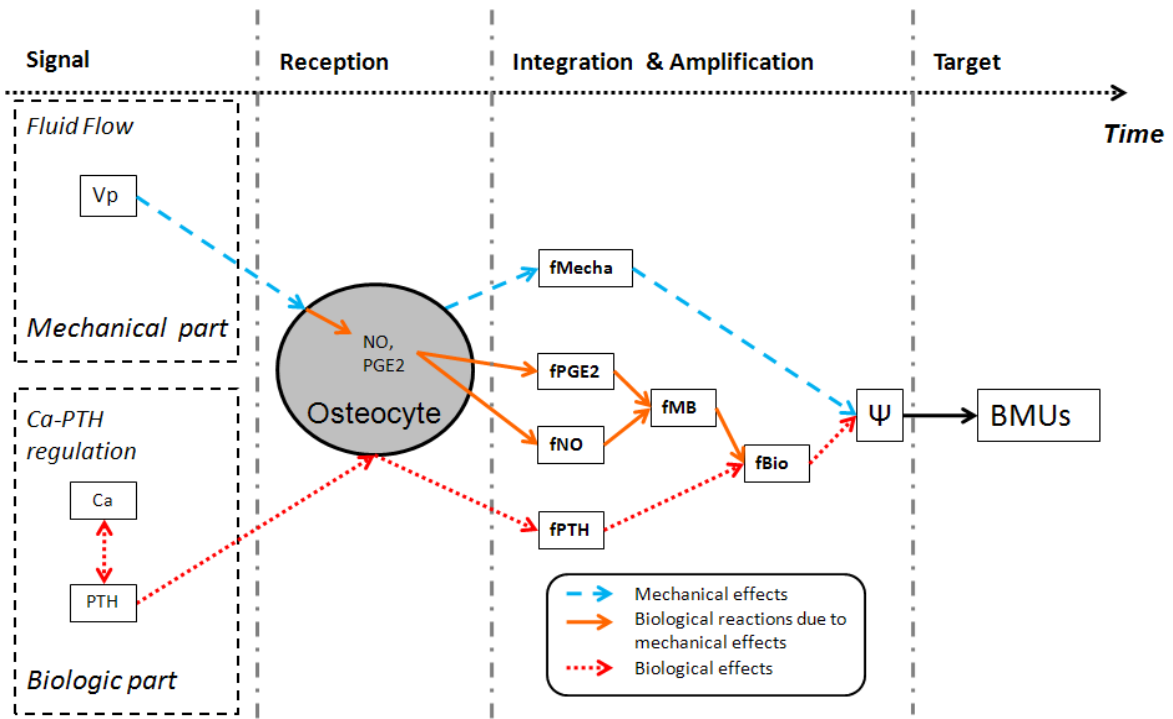

**Figure 1: Block diagram describing the events cascade for bone stimulus with coupled mechanical and biological effects. Dashed arrows represent the mechanical effects and dotted ones represent the biological effects. Solid arrows denote biological reaction due to mechanical stimulation represented by NO & PGE$_2$ production by the osteocyte. Transduction phases (Signal, Reception, Integration & Amplification, Target) are based on Oesterhelt's concept (2010).**

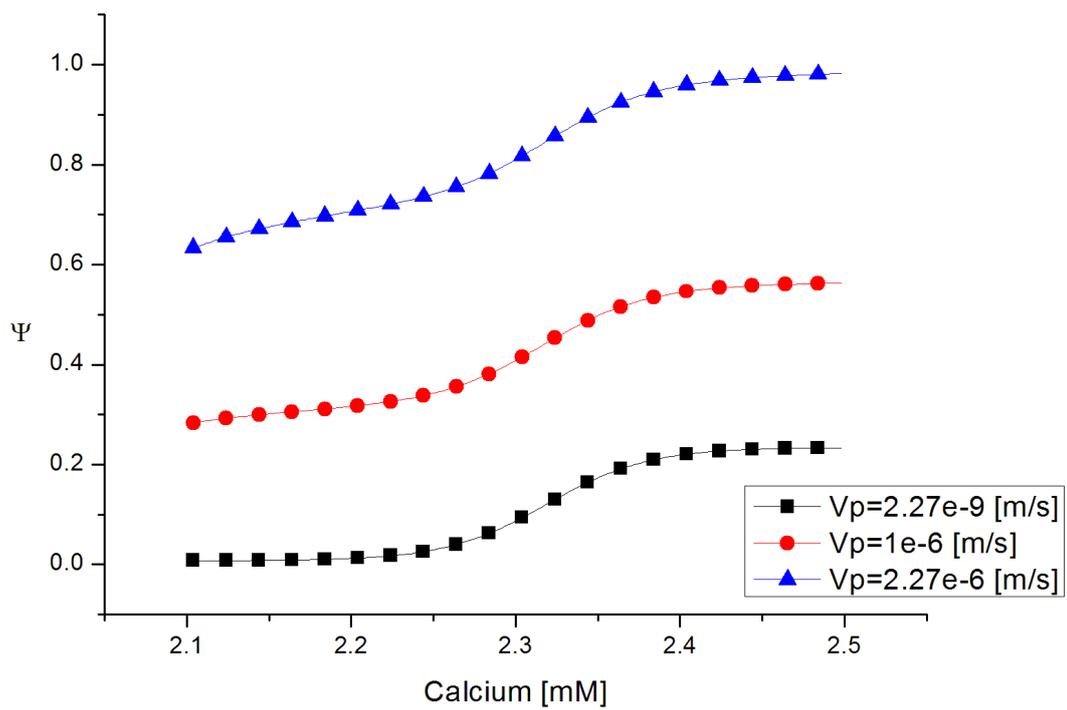

**Figure 2: Mechanobiological stimulus versus calcium concentration for different levels of fluid flow on basal physiological evolution of NO and PGE$_2$.**

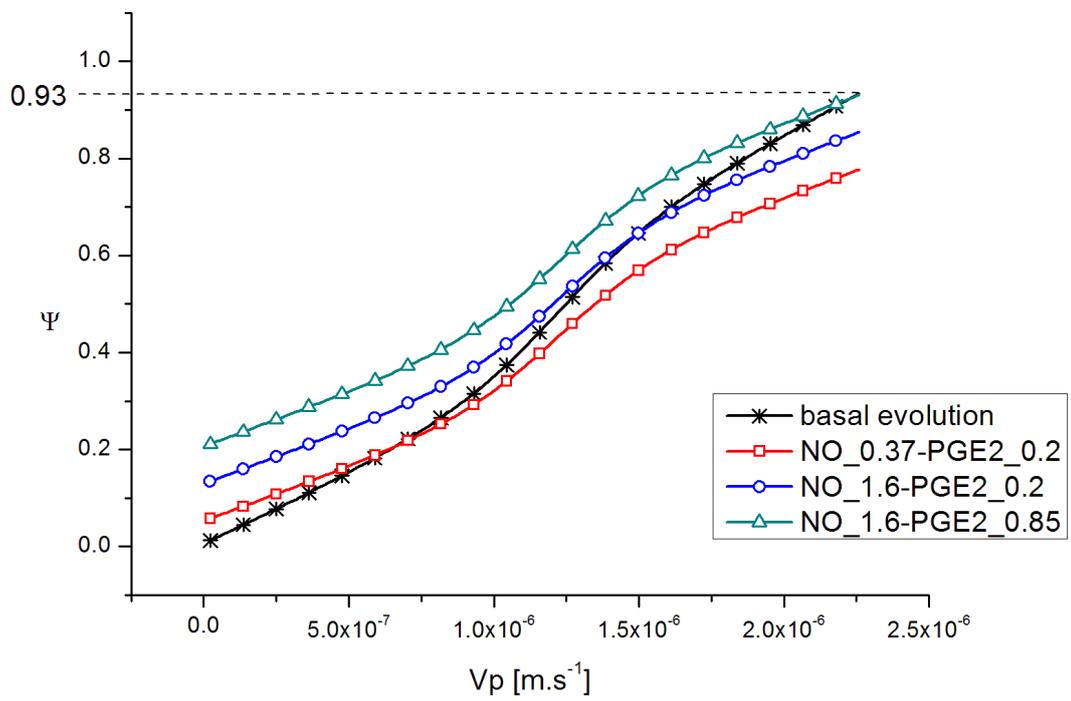

**Figure 3: Predicted stimulus versus fluid velocity for different levels of NO [pM] and PGE$_2$ [pM].**

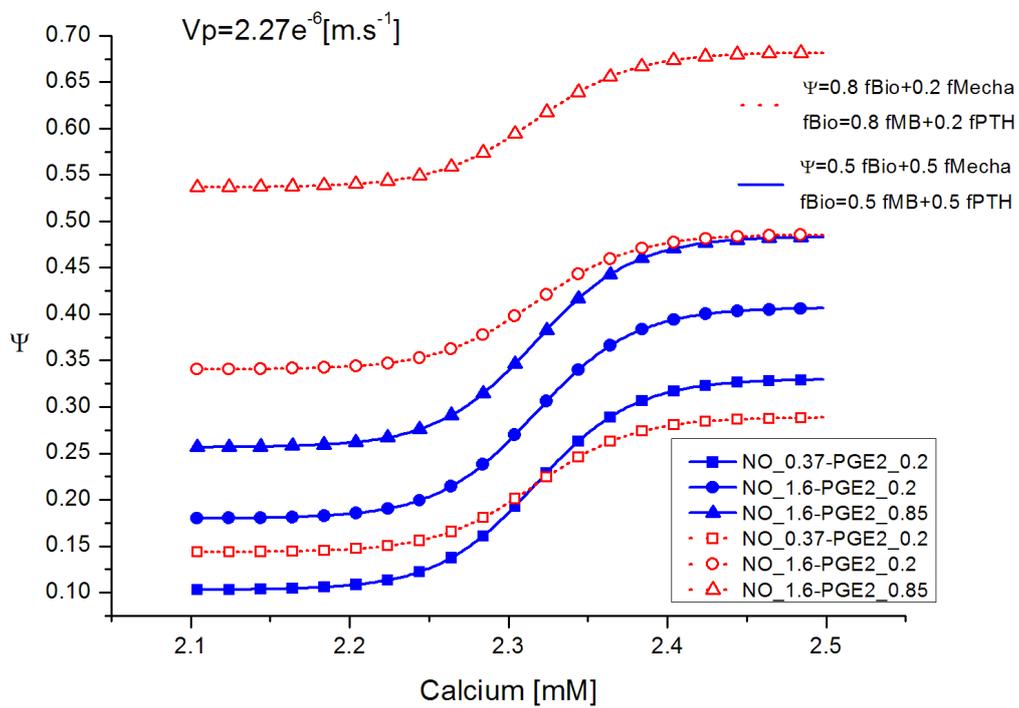

**Figure 4: Predicted stimulus versus calcium concentration at constant fluid flow for different levels of NO [pM], PGE$_2$ [pM] and different configurations of the stimulus Ψ.**

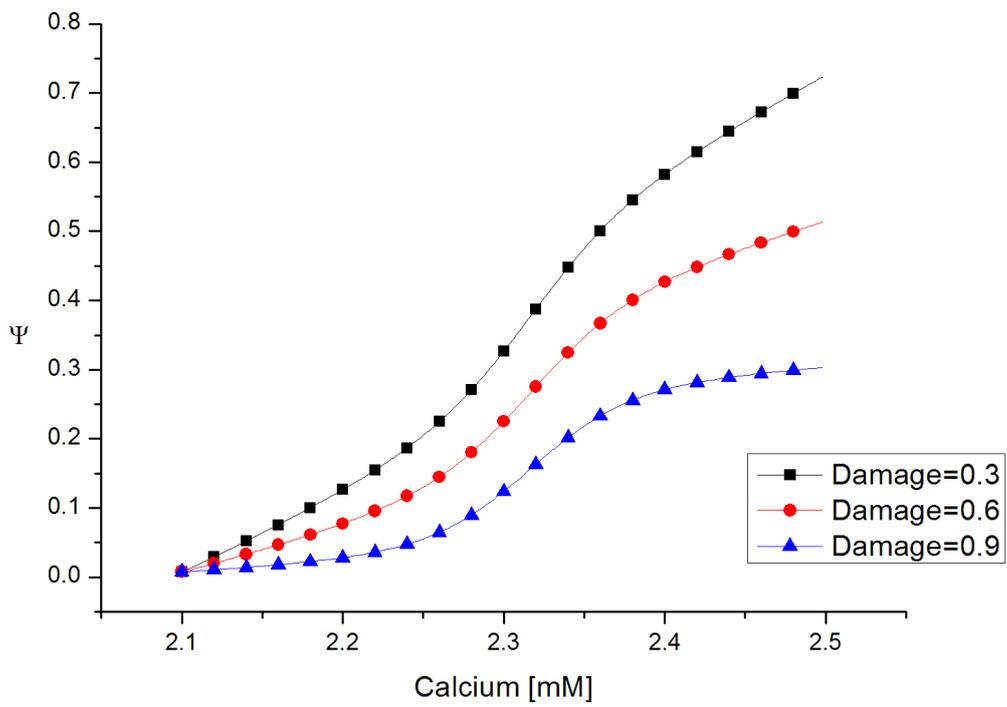

**Figure 5: Predicted mechanobiological stimulus at different levels of canaliculus damages for basal physiological evolution of NO [pM] and PGE$_2$ [pM].**

|            | Signal                     | Reception                  | Integration & Amplification | Target |
|------------|----------------------------|----------------------------|-----------------------------|--------|
| Mechanics  | Fluid flow, Shear stress   | Production of NO & $PGE_2$ | $f_{Mecha}, f_{NO}, f_{PGE2}$ | BMUs   |
| Biology    | Ca                         | PTH regulation             | $f_{PTH}, f_{Bio}$          |        |

**Table 1: Mechanical and biologic components used in the model.**

| | Mechanical part | Ref. | Biological part | Ref. |
|---|---|---|---|---|
| **Signal** | $V_P = -K_P \frac{dP}{dz}(1-D)^\gamma$ (1) | Lemaire et al., 2006 | $x_{Ca} = x_{Ca_0}$ (2) <br><br> $x_{PTH} = \alpha_1 + \frac{\alpha_2}{1+e^{\frac{x_{Ca}-\alpha_3}{\alpha_4}}}$ (3) | Houillier et al., 2000 |
| **Reception** | $V_T = V_P$ (4) | Lemaire et al., 2006 | $\frac{dx_{NO}}{dt} = K_1 f_{V_P} - K_2 x_{NO} + X_{NO}$ (5) <br><br> $\frac{dx_{PGE2}}{dt} = K_3 f_{V_P} - K_4 x_{NO} - K_5 x_{PGE2} + X_{PGE2}$ (6) <br><br> $f_{V_P} = \frac{K_6 V_P K_{14} x_{ocy}}{1+e^{-(K_6 V_P + K_7 x_{ocy})}}$ (7) <br><br> $\frac{dx_{ocy}}{dt} = K_8 f_{PTH}(x_b - X_b) - K_9 f_{V_P}(x_{ocy} - X_{ocy})$ (8) <br><br> $\frac{dx_b}{dt} = K_{10} x_c{}^{g12} x_b{}^{g22} - K_{11} x_b$ (9) <br><br> $\frac{dx_c}{dt} = K_{12} x_c{}^{g11} x_b{}^{g21} - K_{13} x_c$ (10) | Maldonado et al., 2006 <br><br> Maldonado et al., 2006 <br><br> Maldonado et al., 2006 <br><br> Based on Maldonado et al., 2006 <br><br> Komarova et al., 2003 <br><br> Komarova et al., 2003 |
| **Integration & Amplification** | $f_{Mecha} = \frac{V_T}{V_T^M}$ (11) | | $f_{PTH} = \frac{x_{PTH}}{x_{PTH}^M}$ (12) <br><br> $f_{NO} = \frac{x_{NO}}{x_{NO}^M}$ (13) <br><br> $f_{PGE2} = \frac{x_{PGE2}}{x_{PGE2}^M}$ (14) <br><br> $f_{MB} = W_{NO} f_{NO} + W_{PGE2} f_{PGE2}$ (15) <br><br> $f_{Bio} = W_{PTH} f_{PTH} + W_{MB} f_{MB}$ (16) | |
| **Target** | $\Psi = W_{Mecha}\langle f_{Mecha}\rangle + W_{Bio}\langle f_{Bio}\rangle \quad \langle x \rangle = \begin{cases} x, & x > 0 \\ 0, & x \leq 0 \end{cases}$ (17) | | | |

**Table 2: Model's equations**

|  | | Parameter value | Description | Source |
|---|---|---|---|---|
| Fluid Flow | $V_T^M$ | 2.1269e-6 m.s$^{-1}$ | Max. Poiseuille velocity | Goulet et al., 2006 |
| | $K_p$ | 7.5e-20 m².Pa$^{-1}$.s$^{-1}$ | Poiseuille permeability | Gururaja et al., 2005 |
| | $\mu$ | 0.65e-3 Pa.s | Viscosity | Lemaire T. et al., 2008 |
| | $\gamma$ | 2 | Damage exponent | - |
| Calcium | $x_{Ca_0}$ | 2.1 mM | Basal value | Based on Houillier et al., 2009 |
| PTH | $\alpha_1$ | 9.92 pg.ml$^{-1}$ | PTH equ. coef. | Calc. based on Houillier et al., 2009 |
| | $\alpha_2$ | 136.4 pg.ml$^{-1}$ | PTH equ. coef. | Calc. based on Houillier et al., 2009 |
| | $\alpha_3$ | 2.317 mM | PTH equ. coef. | Calc. based on Houillier et al., 2009 |
| | $\alpha_4$ | 3.097e-2 mM | PTH equ. coef. | Calc. based on Houillier et al., 2009 |
| | $x_{PTH}^M$ | 150 pg.ml$^{-1}$ | Max. PTH level | Based on Houillier et al., 2009 |
| NO | $K_1$ | 4e-11 pM.day$^{-1}$ | Release rate | Calc. based on Maldonado et al., 2006 |
| | $K_2$ | 1.e0 day$^{-1}$ | Elimination rate | Maldonado et al., 2006 |
| | $x_{NO}$ | 0.e0 pM.day$^{-1}$ | Ext. admin. rate | Maldonado et al., 2006 |
| | $x_{NO}^M$ | 1.998 pM | Max. level | Calc. based on Maldonado et al., 2006 |
| PGE$_2$ | $K_3$ | 2e-13 pM.day$^{-1}$ | Release rate | Calc. based on Maldonado et al., 2006 |
| | $K_4$ | 1.e-3 day$^{-1}$ | Rate increased by NO | Maldonado et al., 2006 |
| | $K_5$ | 1.e-2 day$^{-1}$ | Elimination rate | Maldonado et al., 2006 |
| | $x_{PGE}$ | 0.e0 pM day$^{-1}$ | Ext. admin. rate | Maldonado et al., 2006 |
| | $x_{PGE}^M$ | 1.06 pM | Max. level | Calc. based on Maldonado et al., 2006 |
| Osteocytes | $K_6$ | 44.e12 s.m$^{-1}$ | Fluid flow influence rate | Calc. based on Maldonado et al., 2006 |
| | $K_7$ | 1.e0 cell$^{-1}$ | Influence rate | Maldonado et al., 2006 |
| | $K_8$ | 1.e-1 day$^{-1}$ | Prod. rate | Maldonado et al., 2006 |
| | $K_9$ | 2.003e-11 day$^{-1}$ | Degradation rate | Calc. based on Maldonado et al., 2006 |
| | $K_{14}$ | 1.e0 cell$^{-1}$ | Influence rate | Calc. based on Maldonado et al., 2006 |
| | $x_{ocy}$ | 500 cells | Ref. value | Komarova et al., 2003 – Parfitt 1977 |
| Osteoblasts | $x_b$ | 200 cells | Ref. value | Komarova et al., 2003 – Parfitt 1977 |
| | $K_{10}$ | 3.e0 cell.day$^{-1}$ | Prod. rate | Komarova et al., 2003 |
| | $K_{11}$ | 2.e-1 cell.day$^{-1}$ | Elimination rate | Komarova et al., 2003 |
| | $g_{12}$ | 1.e0 | Paracrine factor | Komarova et al., 2003 |
| | $g_{22}$ | 0.e0 | Autocrine factor | Komarova et al., 2003 |
| Osteoclasts | $x_c$ | 20 cells | Ref. value | Komarova et al., 2003 – Parfitt 1977 |
| | $K_{12}$ | 4.e0 cell.day$^{-1}$ | Prod. rate | Komarova et al., 2003 |
| | $K_{13}$ | 2.e-2 cell.day$^{-11}$ | Elimination rate | Komarova et al., 2003 |
| | $g_{11}$ | 0.5e0 | Autocrine factor | Komarova et al., 2003 |
| | $g_{21}$ | -0.5e0 | Paracrine factor | Komarova et al., 2003 |

**Table 3: Model's parameters**